# Study on the Time Domain Precision Evolution Mechanism of CNC Machine Tool Feed Systems Based on Acceleration and Deceleration Capability Indicator

Xuesong Wang[1], Yi Zhou[2], Dongsheng Zhang[1]*

[1]Department of Mechanical Engineering, Xi'an Jiaotong University, Xi'an 710064 China

[2] Beijing Institute of Space Launch Technology, Beijing 100000 China

Corresponding author: Dongsheng Zhang (zds@xjtu.edu.cn)

## Abstract

The escalating demand for high-speed and high-precision machining in machine tool feed system has brought to the forefront the challenge of its design method. Currently, existing methodologies struggle to ascertain compliance with dynamic performance requirements during the design phase, often resulting in either excessive or insufficient design. Therefore, there is an urgent need for research focused on feed system design methods that directly address time domain dynamic precision. The dynamic precision of the feed system is influenced by the motor, mechanical structure, motion processes, and control system. However, existing studies on the impact mechanisms of electromechanical matching on feed system precision often overlook the roles of control and motion processes. This paper innovatively proposes the need to consider the coupling effects among subsystems, directing the optimization design of CNC machine tool feed systems towards time domain dynamic precision. Furthermore, it introduces acceleration and deceleration capability as a key indicator of electromechanical matching. Following the decoupling of control system parameters, this study elucidates the influence mechanisms of electromechanical matching on the overall dynamic performance of the feed system under various motion processes. This research offers a novel design philosophy and theoretical foundation for the optimization of CNC machine tool feed systems.

Keywords: Time domain precision; Precision oriented approach; High-speed machine tools; Machine tool dynamics; Acceleration and deceleration capacity;

## 1 Introduction

The feed system is a critical component in determining the machining precision of CNC machine tools [1–4]. The industry focuses on dynamic performance indicators of CNC machine tool feed systems, primarily including feed speed, positioning accuracy, and repeatability as time domain metrics. Current design methodologies typically employ lightweight and structural optimization to shift the natural frequency away from operational frequencies, thereby enhancing dynamic precision [5–11]. However, frequency-domain analysis methods are qualitative, and the relationship between natural frequency indicators and time-domain precision metrics is not one-to-one. This frequency-based design approach makes it challenging to accurately assess compliance with dynamic performance requirements during the design phase, often leading to either excessive or insufficient designs. Therefore, it is essential to consider positioning accuracy $\Delta e(t)$ and velocity accuracy $\Delta \dot{e}(t)$ as time domain precision indicators, investigating design methods for CNC machine tool feed systems that directly address time domain dynamic precision.

Positioning accuracy $\Delta e(t)$ and velocity accuracy $\Delta \dot{e}(t)$ are fundamentally kinematic indicators, with acceleration as their core component. Acceleration, velocity, and displacement are interrelated through time integration. Acceleration serves as a crucial link between kinematics and the physical world [12,13]. The electromechanical matching within the physical realm dictates the upper limits of the feed system's acceleration and deceleration performance, subsequently influencing its motion characteristics and dynamic precision. Currently, design approaches typically utilize the load inertia ratio as an indicator for electromechanical matching [14,15]. However, this method does not intuitively reflect the motion characteristics of the feed system. Therefore, this paper exploratively proposes using the acceleration and deceleration capability of the feed system as a key indicator for examining the electromechanical matching relationship. The expression for acceleration and deceleration capability $C_{AD}$ is given in Equation (1), where $T_{\max}$ represents the maximum torque of the motor, $J_m$ denotes the motor's inertia, $J_L$ signifies the mechanical system's inertia, and $R_r$ is the mechanical system's transmission ratio.

$$C_{AD} = \frac{T_{\max}}{J_m + J_L} R_r \tag{1}$$

Electromechanical matching determines the system's acceleration and deceleration capabilities, which are critical factors influencing the time domain precision of the feed system. However, analysis of the block diagram of the feed system, as shown in Figure 1, indicates that its time domain dynamic precision is influenced by the integrated effects of four subsystems: motion processes, control systems, drive motors, and mechanical systems [16], as expressed in Equation (2). Therefore, when conducting electromechanical matching design, it is essential to consider the impacts of control and motion processes [17]. Scholars have initiated research in this area. For instance, Weck et al. [18] demonstrated that the tuning of control parameters should account for mechanical characteristics. Changes in motion processes affect factors such as friction and interface properties [19–21], subsequently influencing the dynamic performance of the feed system. Research by Zhang et al. [22,23] found that the natural frequency of the feed system varies with changes in motion process. However, a systematic investigation into how to comprehensively consider the effects of control and motion processes, and to directly study the mechanisms by which electromechanical matching influences time domain precision, is still lacking.

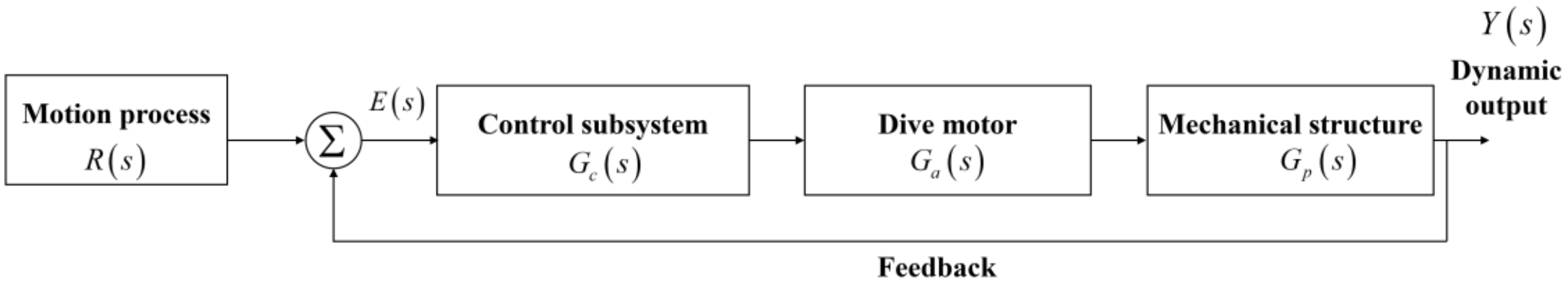

Figure 1 Subsystem block diagram of the feed system

$$\Delta e(t) = \mathcal{L}^{-1}\left[E(s)\right] = \mathcal{L}^{-1}\left[\frac{R(s)}{1+G_c(s)G_a(s)G_p(s)}\right] \tag{2}$$

In summary, existing design methodologies struggle to ascertain whether they meet time domain performance requirements during the design phase, failing to establish direct dynamic design approaches focused on temporal precision. Moreover, systematic studies incorporating the effects of control and motion processes in electromechanical matching design are lacking. To address this issue, this paper innovatively proposes using time domain dynamic precision as a direct design indicator, introducing acceleration and deceleration capability as a key indicator to characterize the electromechanical matching relationship. This approach comprehensively considers the influences of control and motion processes while directly investigating the mechanisms by which electromechanical matching affects temporal precision. This research advocates for an optimization design of CNC machine tool feed systems that are explicitly oriented towards temporal dynamic precision, revealing the impact mechanisms of electromechanical matching on the overall dynamic performance of the feed system under various motion processes after decoupling control system parameters. This work provides a novel design philosophy and theoretical foundation for the optimization of CNC machine tool feed systems.

The organization of this paper is as follows: First, an integrated model of the feed system incorporating "electromechanical, control, and motion processes" is established. Next, the influence of the control system is isolated using intelligent optimization algorithms. Subsequently, the mechanisms by which variations in motion processes and acceleration/deceleration capabilities affect the performance of the feed system are investigated. Finally, an experimental platform is constructed for validation.

## 2 Integrated Modeling of Electromechanical Control and Motion Process in Feed System

The feed system is a sophisticated integrated system, comprising mechanical structure, motor, and control subsystems that collectively affect its dynamic performance. To analyze the impact of various subsystems on the feed system's dynamic performance, it is crucial to construct a feed system model that accurately reflects its actual working characteristics. In this paper, an integrated feed system dynamics model that incorporates motor, mechanical structure, control system, and motion process is used for analysis.

In the feed system, the servo motor serves as the power unit responsible for driving the motion of the mechanical structure. The mechanical structure outputs the speed and displacement of the table. These outputs are then fed back into the control system. The control system can convert the error between actual motion and planned motion into torque commands to propel the motor. Then the closed loop and joint simulation of the integrated feed system dynamics model is formed.

### 2.1 Modeling of motion processes

The motion process of the feed system is carried out using a trapezoidal speed planning curve, as illustrated in Figure 2. This curve encompasses three distinct phases: the acceleration phase, the uniform speed phase, and the deceleration phase.

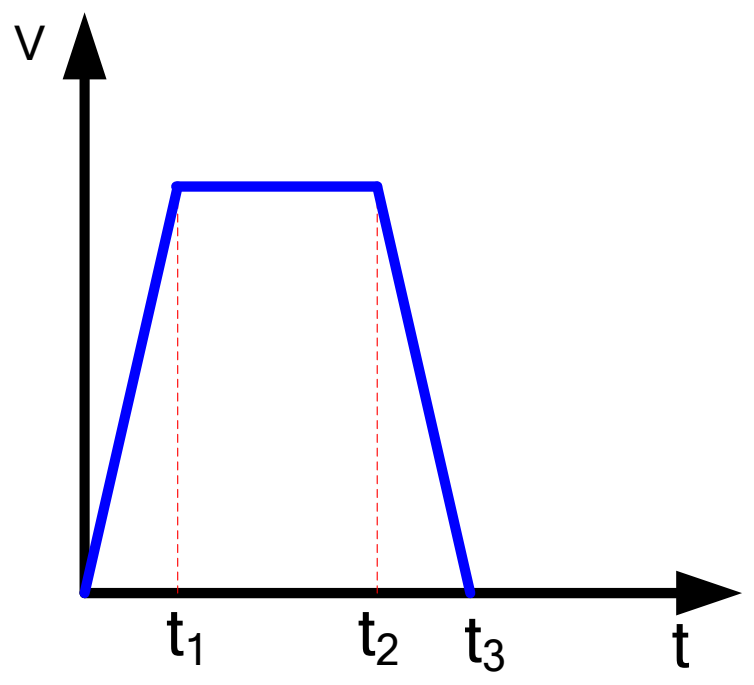

Figure 2 Trapezoidal speed planning curve

The specific calculation process is as follows:

(1) Acceleration phase：

$$v(t)=at,\quad s_a=\frac{1}{2}at^2 \tag{3}$$

(2) Uniform phase：

$$v(t)=v,\quad s_u=v(t_2-t_1) \tag{4}$$

(3) Deceleration phase：

$$v(t)=v-a(t_3-t_2),\quad s_d=\frac{1}{2}a(t_3-t_2)^2 \tag{5}$$

Where $a$ is acceleration value; $v$ is maximum speed value; $s_a$ is acceleration distance; $s_u$ is constant velocity distance; $s_d$ is deceleration distance.

This paper will change the maximum speed $v$ and acceleration $a$ of the feed system. This variation will enable the feed system to reciprocate at different motion processes, facilitating an investigation into the performance changes of the feed system under diverse motion processes.

### 2.2 Modeling Servo Control Systems

Moving forward, the focus shifts to constructing the control system for the feed system. Typically, servo feed system controllers adopt a multi-loop PID controller, comprising position, velocity, and current loops. This configuration aims to bolster the system's anti-interference capability and dynamic tracking performance[24]. For this study, a two-loop PID and velocity feedforward control system, as illustrated in Figure 3, is employed.

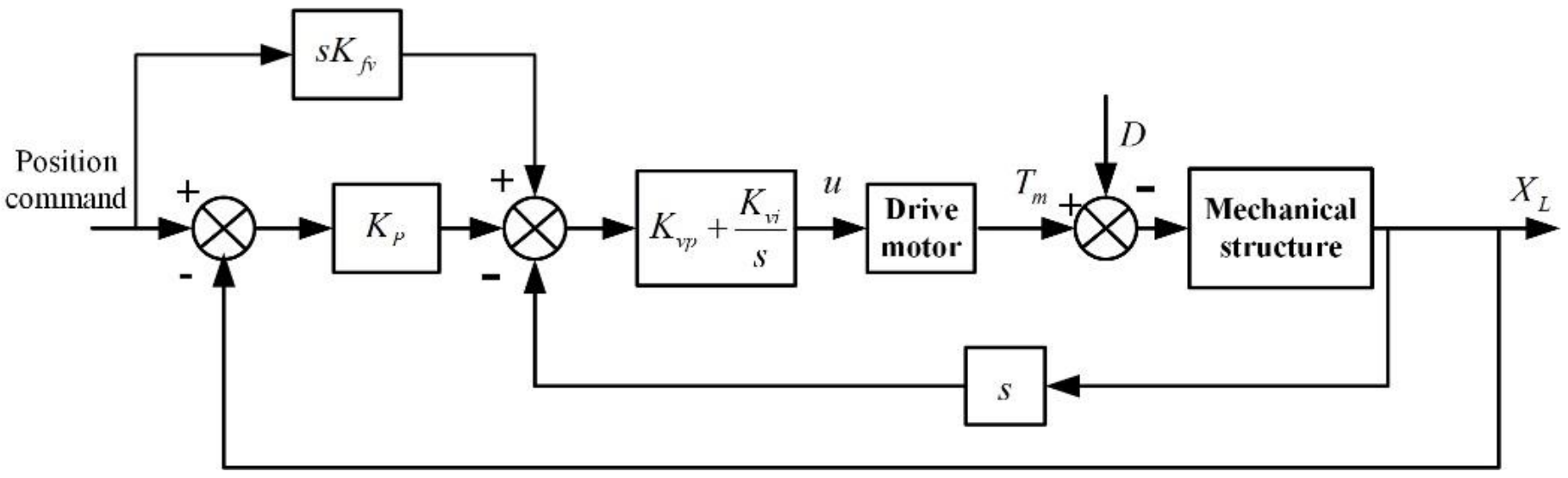

Figure 3 Controller Block Diagram

In the provided notation, $K_{fv}$ represents the speed feedforward proportional coefficient; $K_p$ represents the position loop proportional coefficient; $K_{vp}$ represents the speed loop proportional coefficient; $K_{vi}$ represents the speed loop integral coefficient; $D$ represents the load torque; $T_m$ represents the motor output torque.

In this system, the position loop employs proportional control to achieve precise position control, ensuring dynamic tracking performance. The speed loop utilizes a PI controller to enhance the system's resistance to disturbances and suppress speed fluctuations. The current loop, having a significantly larger bandwidth than the speed or position controllers in the servo system, is considered equivalent to the torque constant in this paper. Subsequently, an optimization algorithm will be applied to fine-tune the four control parameters, enhancing the dynamic performance of the system.

### 2.3 Modeling of Electromechanical Systems

In this paper, the mechanical model of the feed system is created using SolidWorks. Recognizing the inherent complexity of the actual mechanical model of the servo feed system, certain simplifications are applied to maintain computational efficiency. Figure 4 illustrates the simplified mechanical structure of the feed system. Detailed mechanical characteristics are outlined in Table 1.

To facilitate joint simulation with the motor, this paper exports the SolidWorks 3D model into the MATLAB environment to construct a dynamics model, as depicted in

Figure 5. By selecting different types of motors, the acceleration and deceleration capacity of the feed system can be changed.

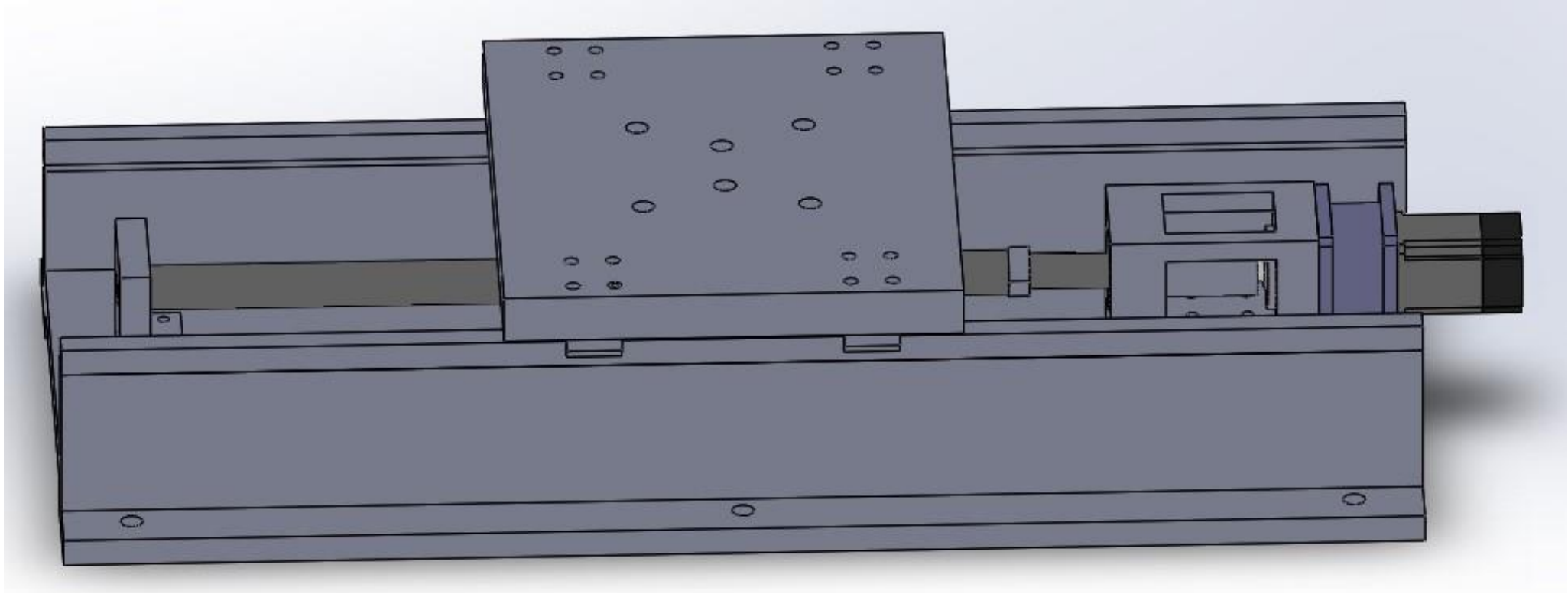

Figure 4 Three-dimensional model of the feed system

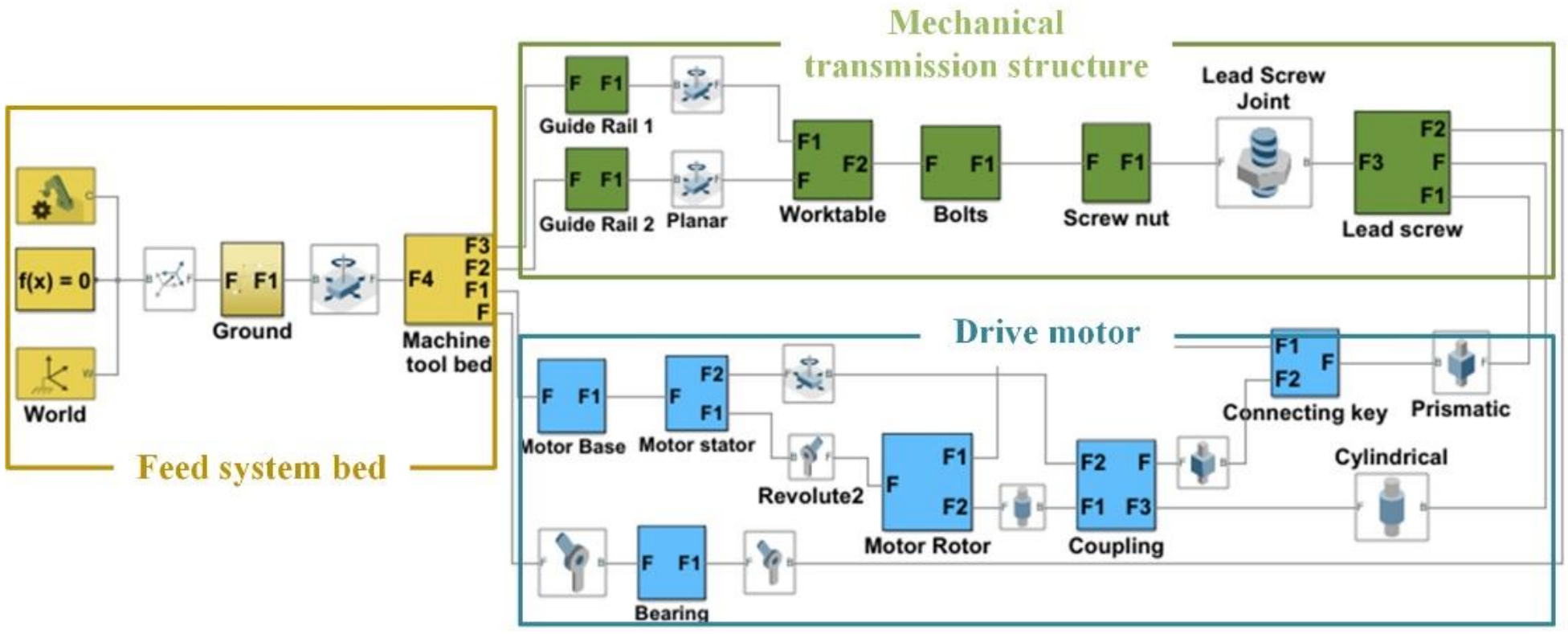

Figure 5 Virtual prototype mechanical model

Table 1 Characteristic parameters of mechanical structures

| Variables | Meaning | Values | Unit |
|---|---|---|---|
| $K$ | Lead screw stiffness | 612 | $\mathrm{N \cdot m \cdot rad^{-1}}$ |
| $J_L$ | Load inertia | 45.5 | $\mathrm{kg \cdot cm^2}$ |
| $B$ | Damping | 0.0288 | $\mathrm{kg \cdot m^2 \cdot s^{-1}}$ |
| $R$ | The transmission coefficient of lead screw | $10/2\pi$ | $\mathrm{mm \cdot rad^{-1}}$ |

## 3 Isolating the effect of the control system on the dynamic performance of the feed system

This paper aims to investigate the electromechanical matching relationship while considering the influences of control parameters and motion processes. Simultaneously studying the effects of control, motion processes, motors, and mechanical structures on feed system performance involves numerous design parameter variables, making analysis challenging. Therefore, this study introduces the concept of control decoupling, optimizing control parameters through algorithms to achieve optimal feed system performance. This approach isolates the influence of the control system on the overall dynamic performance of the feed system, enabling a focused investigation of the mechanisms by which different motion processes and acceleration/deceleration capabilities affect the system.

This paper adopts the control loop illustrated in Figure 3, comprising the position loop, velocity loop, and velocity feedforward. As a result, the overall control parameters to be optimized include $K_P$, $K_{vi}$, $K_{vp}$ and $K_{fv}$. The optimization objectives in this study focus on the position accuracy and velocity accuracy of the system. Position accuracy is assessed using the system's following error as the evaluation index, while velocity accuracy considers the system's speed error and velocity fluctuation as evaluation indices[25]. The overall comprehensive optimization objectives are shown in equation (6).

$$W = 0.5 * \max(Err_p) + 0.25 * \max(Err_v) + 0.25 * Vars_v \quad (6)$$

Where, $Err_p$, $Err_v$ and $Vars_v$ correspond to the following error, speed error, and speed fluctuation of the system, respectively.

For different motion processes and ADC, the optimization of control parameters is conducted with the position and speed accuracy of the system as the target. To ensure the credibility of the optimization results and achieve the best dynamic performance, this paper employs different kinds of optimization algorithms for cross-validation. In this study, genetic algorithms(GA), fireworks algorithms(FA), and an Migration genetic algorithm(MGA) were employed for six optimization iterations, with results illustrated in Figure 6. The horizontal axis represents the number of optimization iterations, while the vertical axis shows the maximum position error (left) and velocity error (right). To facilitate a more intuitive analysis of the performance of different parameter optimization algorithms, we quantified the fluctuations in Figure 6 using variance, as presented in Table 2.

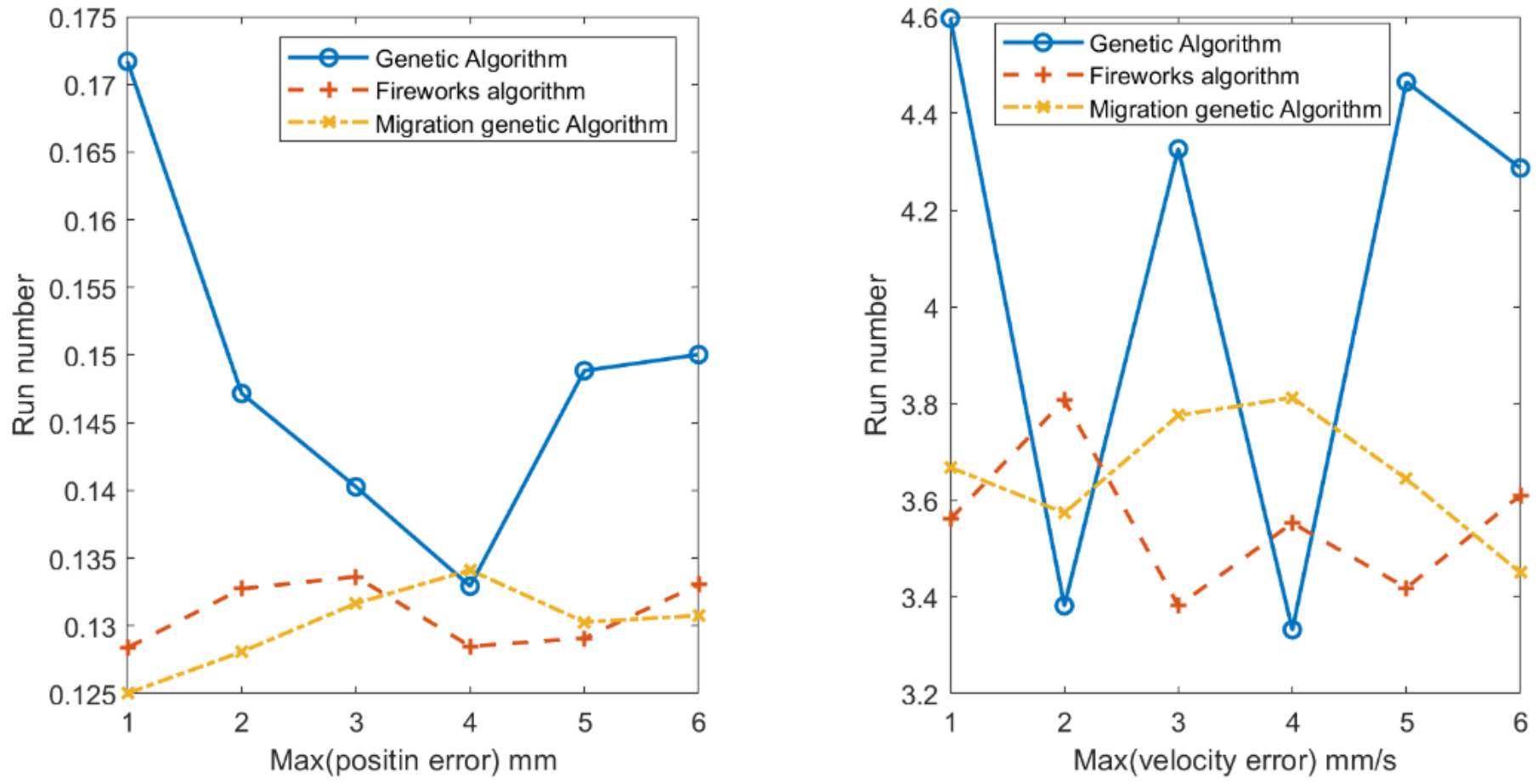

Figure 6 Results of different optimization algorithms

Table 2 Results of different optimization algorithms

| | GA | FWA | MGA |
|---|---|---|---|
| Variance (maximum position error) | $1.71\times10^{-4}$ | $6.17\times10^{-6}$ | $9.78\times10^{-6}$ |
| Variance (maximum velocity error) | 0.3129 | 0.0231 | 0.0176 |

The results shown in Figure 6 and Table 2 indicate that the performance of the GA is significantly inferior to FWA and MGA. Both FWA and MGA demonstrate higher stability and better optimization results. To ensure the effectiveness of control parameter decoupling and to optimize the performance of the feed system, subsequent decoupling will involve mutual validation between FWA and MGA. This approach aims to ensure the reliability of the optimization results.

## 4 Analysis of the influence of the acceleration and deceleration capacity of the feed system under different motion

**processes**

By varying the system's motion process parameters and electromechanical parameters, an analysis of the influence of the electromechanical system and motion process of the feed system is performed.

A reciprocating movement of the workbench, covering a distance of 200 mm, is conducted. By altering acceleration and maximum velocity, the motion processes can be changed. The motion process parameters are shown in Table 3. By selecting different types of Inovance motors, the acceleration and deceleration capacity of the feed system can be changed. The selected motor's parameters are outlined in

Table 4. Through the construction of various motion processes and servo motor sequences, this study analyzes the impact of acceleration and deceleration capacity on system performance under different motion processes.

Table 3 Motion process parameters

| Speed/ $mm \cdot s^{-1}$ | 100 | 100 | 100 | 200 | 200 | 200 | 400 | 400 | 400 |
|---|---|---|---|---|---|---|---|---|---|
| Acceleration/ $m \cdot s^{-2}$ | 1 | 2 | 5 | 1 | 2 | 5 | 1 | 2 | 5 |

Table 4 Servo motor related parameters

| Serial number | Motor Model | Maximum Torque $T_{max}$ / $N \cdot m$ | Rotor inertia $J_m$ / $kg \cdot cm^2$ | ADC $\frac{T_{max} R}{J_m + J_L}$ |
|---|---|---|---|---|
| 1 | ISMH3-44C15CD | 71.1 | 88.9 | 8.4 |
| 2 | ISMH3-29C15CD | 37.2 | 55 | 5.9 |
| 3 | ISMH3-18C15CD | 28.75 | 25.5 | 6.4 |
| 4 | ISMH3-13C15CD | 20.85 | 19.3 | 5.1 |
| 5 | ISMH3- | 13.5 | 13 | 3.6 |
| 6 | 85B15CD<br>1MH3-50B15CB | 9.6 | 11.01 | 2.7 |

Following the optimization of servo parameters, the paper proceeds to analyze the impact of acceleration and deceleration capacity on system performance under different motion processes. This study employs the position error, velocity error, and velocity fluctuation of the system to comprehensively evaluate system accuracy [26]. The simulation results illustrate the corresponding changes in system performance $W$ for different Inovance motors under various processes, as depicted in Figure 7.

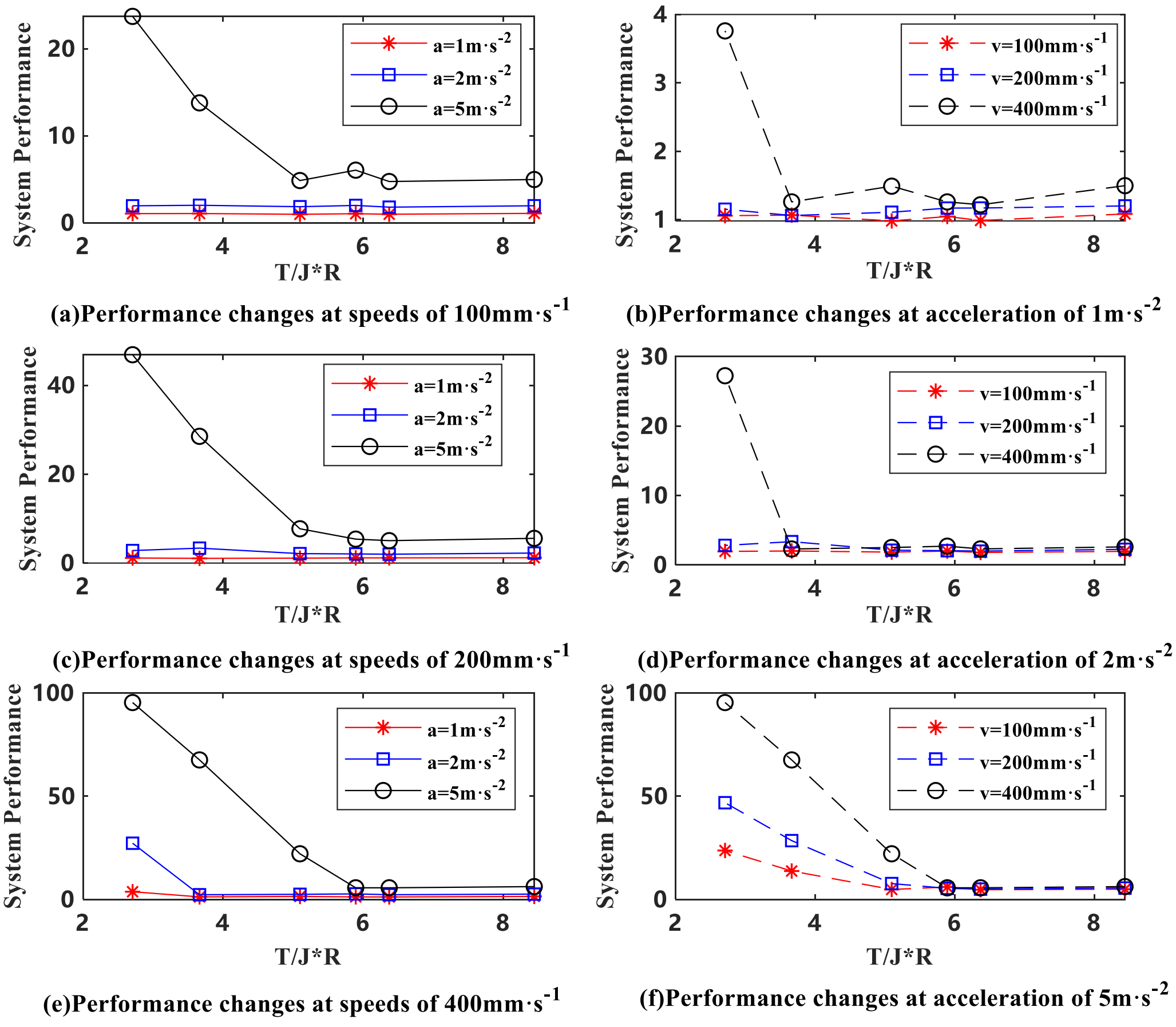

(a)Performance changes at speeds of 100mm·s$^{-1}$

(b)Performance changes at acceleration of 1m·s$^{-2}$

(c)Performance changes at speeds of 200mm·s$^{-1}$

(d)Performance changes at acceleration of 2m·s$^{-2}$

(e)Performance changes at speeds of 400mm·s$^{-1}$

(f)Performance changes at acceleration of 5m·s$^{-2}$

Figure 7 Variation of system performance of different motors at different speeds and accelerations

Figure 7 illustrates the variation in system performance at different speeds and accelerations. The horizontal axis represents ADC. The vertical axis represents the system's comprehensive evaluation index $W$. A smaller $W$ value indicates better system performance. It can be observed that under different motion processes, as the acceleration and deceleration capacity of the system increases, system performance gradually improves and eventually stabilizes. This suggests that when the overall acceleration and deceleration capacity of the system is below the actual motion process demand, system performance improves with increasing acceleration and deceleration capacity. However, when the overall acceleration and deceleration capacity of the system surpasses the actual motion process demand, system performance tends to remain stable.

The presented results highlight that the acceleration and deceleration capacity of the system is a crucial factor influencing system performance. In ideal conditions, increasing the overall acceleration and deceleration capacity of the system can enhance its performance. However, in real-world working processes, uncertainties may be present, necessitating consideration of the system's relative stability.

This paper evaluates the stability in terms of the phase margin and amplitude margin,. A larger phase margin and amplitude margin indicate higher stability, while a larger relative resonance peak corresponds to a larger overshoot in the step response of the system, indicating lower stability. The amplitude margin and phase margin of the system are employed as constraints to ensure that the system meets the following criteria: the phase margin is greater than 30, the amplitude margin is greater than 6 dB [27].

$$\begin{cases} A_m > A_{\min} = 6dB \\ P_m > P_{\min} = 30^o \end{cases} \tag{7}$$

By maintaining the relative stability of the system within a threshold range, the control parameters are optimized again. This optimization is based on the relative stability of the system to achieve the decoupling of control parameters. The system performance, considering constraints to ensure the relative stability of the system, is analyzed, and the results are presented in Figure 8.

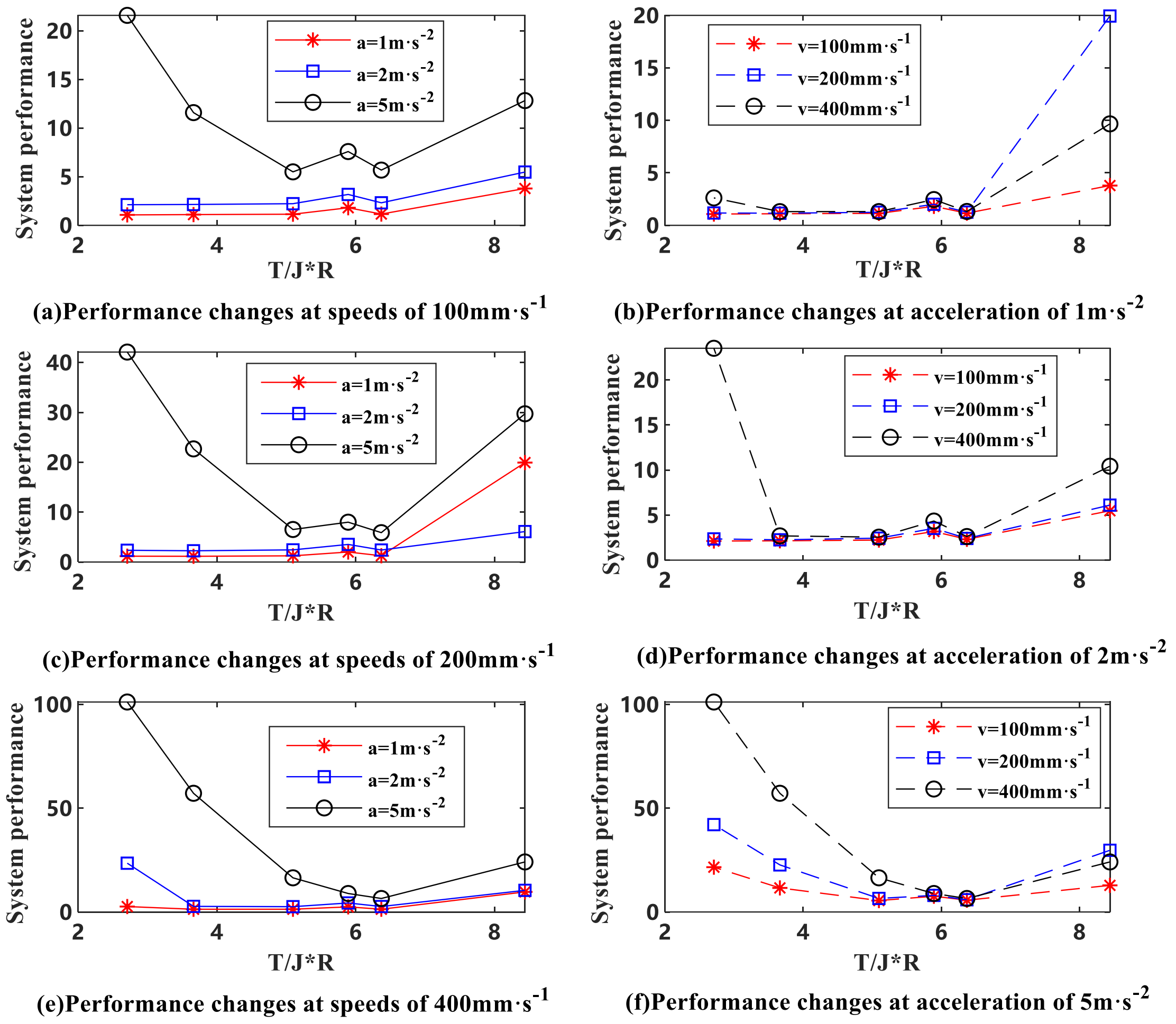

Figure 8 Variation in system performance corresponding to different motors considering stability

The change in the overall performance of the system for different acceleration and deceleration capacities is obtained under the premise of relative stability constraints, where a smaller value on the vertical axis indicates better system performance. Comparing Figure 8 with Figure 7, the variation is as follows: as the acceleration and deceleration capacity of the system increases, the system performance initially improves and then decreases. This suggests that an increase in acceleration and deceleration capacity will have a certain impact on the stability of the system. When the overall acceleration and deceleration capacity of the system meets the actual motion demand, the system performance decreases as the acceleration and deceleration capacity continues to increase.

For a more intuitive analysis of the relationship between acceleration and deceleration capacity and system stability, the relative change in performance between considering the stability constraint and not considering the stability constraint is provided, as shown in Figure 9.

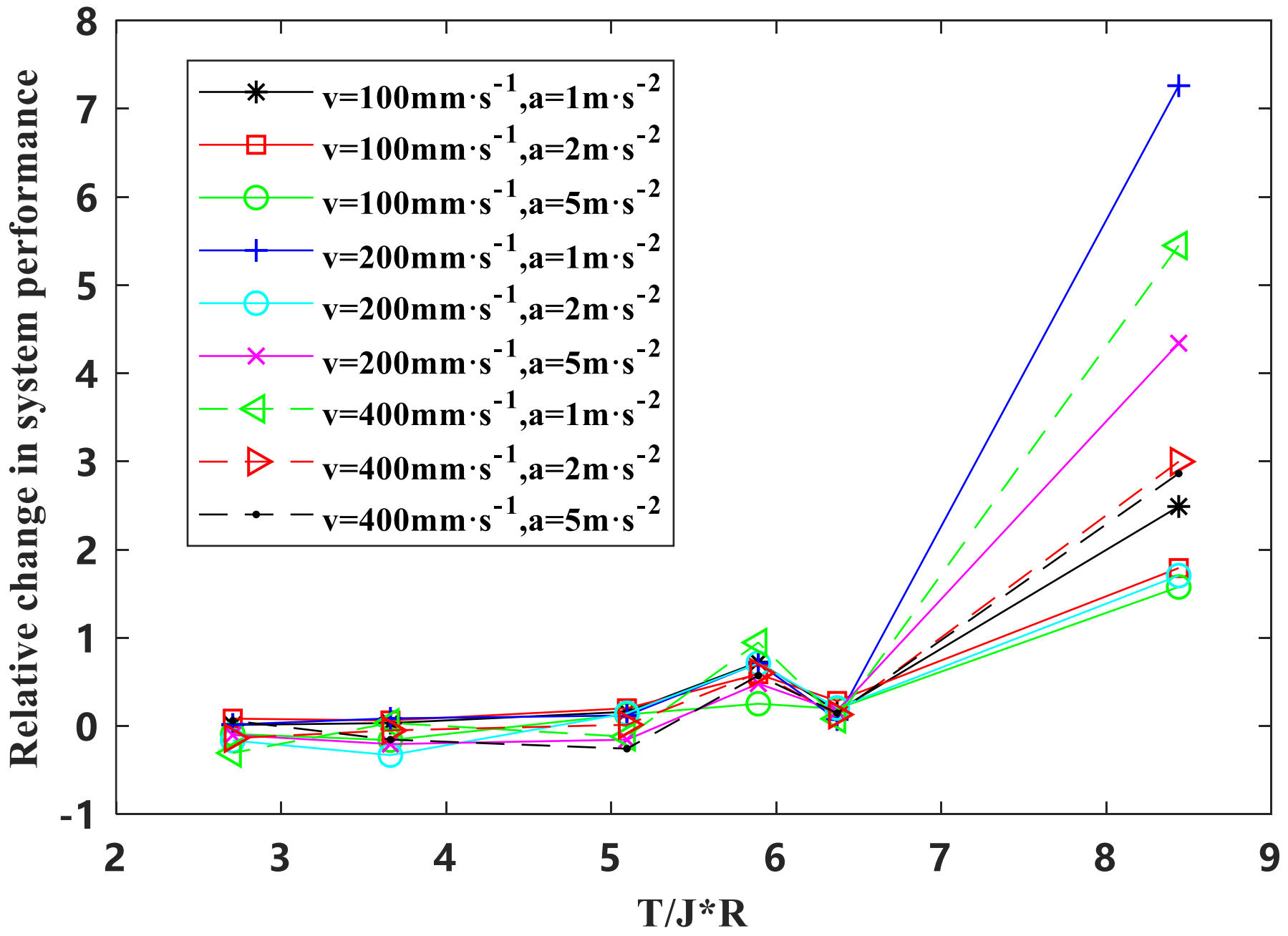

Figure 9 Change in performance with stability considered compared to the case without stability considered

As depicted in the figure, the horizontal and vertical coordinates represent the system acceleration capacity and the relative change in performance, respectively. The relative change in system performance is expressed as $[(w_{stable} - W_{unstable})/W_{unstable}]$. It can be observed that the overall relative change in system performance is greater than 0, indicating a decrease in system performance under the premise of ensuring system stability. Furthermore, as the acceleration and deceleration capacity increases, the relative change in performance gradually increases. In other words, the magnitude of the decrease in system performance increases, suggesting that the acceleration and deceleration capacity of the system have a negative impact on system stability.

In summary, on one hand, an increase in the acceleration capacity of the system enhances its ability to follow motion commands, leading to improved dynamic performance under ideal operating conditions. On the other hand, this increase in acceleration capacity can adversely affect the stability of the system. Therefore, to ensure both dynamic performance and stability, it is crucial for the overall acceleration capacity of the system to meet actual motion requirements without pursuing excessive acceleration and deceleration capacity.

## 5 Experimental validation

To verify the effect of the acceleration and deceleration capacity on the system dynamic performance, an experimental platform was constructed. This platform comprised Inovance servo motors, servo driver, mechanical transmission system, and load, as illustrated in Figure 10.

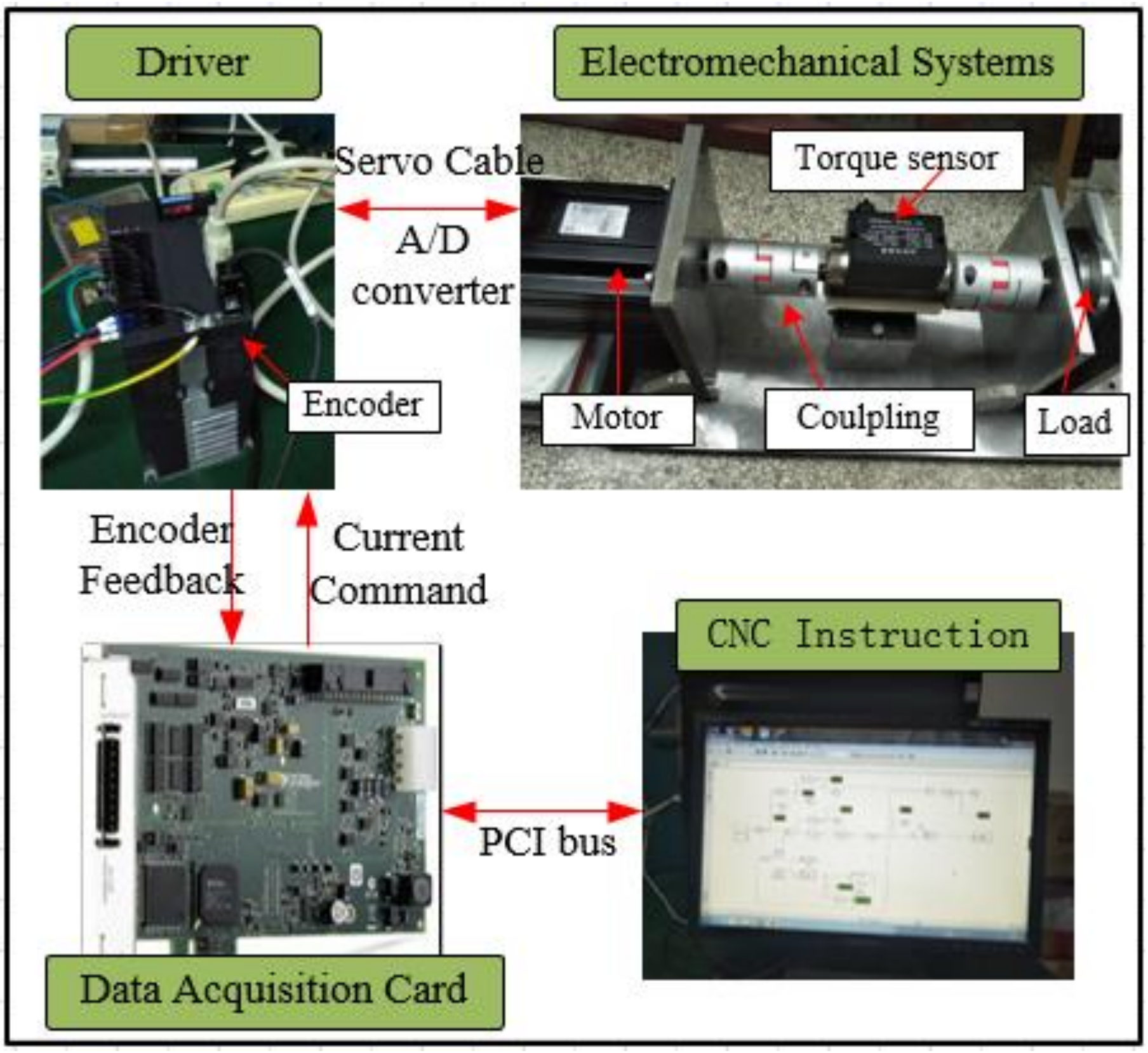

Figure 10 Experimental platform

The motor driver is connected to the computer through the NI PCIe-6321 board for signal transmission. The signal is collected through the motor encoder, serving as the feedback mechanism. The specific software operation involves using MATLAB to drive the NI board. Simulink's Analog Output block is employed to send target commands to drive the load rotation. Simulink's Encoder Input block processes the collected signal from the encoder and divides it by the motor encoder's resolution of 2500 to form a feedback loop.

Combined with the simulation analysis results and the actual working conditions, three groups of Inovance motors are selected for the experimental platform, and the specific parameters are shown in Table 5.

Table 5 Specific parameters of the experimental motor

| Serial number | Motor Model | Maximum Torque $T_{\max}$ / $\mathrm{N \cdot m}$ | Rotor inertia $J_m$ / $\mathrm{kg \cdot cm^2}$ | Load inertia $J_L$ / $\mathrm{kg \cdot cm^2}$ | Inertia Ratio $r_l$ | ADC $\frac{T_{\max} R}{J_m + J_L}$ |
|---|---|---|---|---|---|---|
| A | ISMH3-29C15CD | 37.2 | 55 | 48 | 0.8 | 5.7 |
| B | ISMH3-18C15CD | 28.75 | 25.5 | 48 | 1.9 | 6.2 |
| C | ISMH3-85B15CD | 13.5 | 13 | 48 | 3.7 | 3.5 |

Using the three groups of motors for experiments, the fixed load, i.e., the working turntable, remains unchanged, and the drive motor is changed three times. The motion process is consistent with the process parameters in the simulation. The control parameters are automatically adjusted through the Inovance motor driver so that each group of experiments works in the optimum state. To facilitate the experiment, a turntable was used as the load, and thus, the following error, speed error, and speed fluctuation corresponded to the angular error, angular speed error, and angular speed fluctuation, respectively. The experimental results under different motion processes are depicted in Figure 11 to Figure 16.

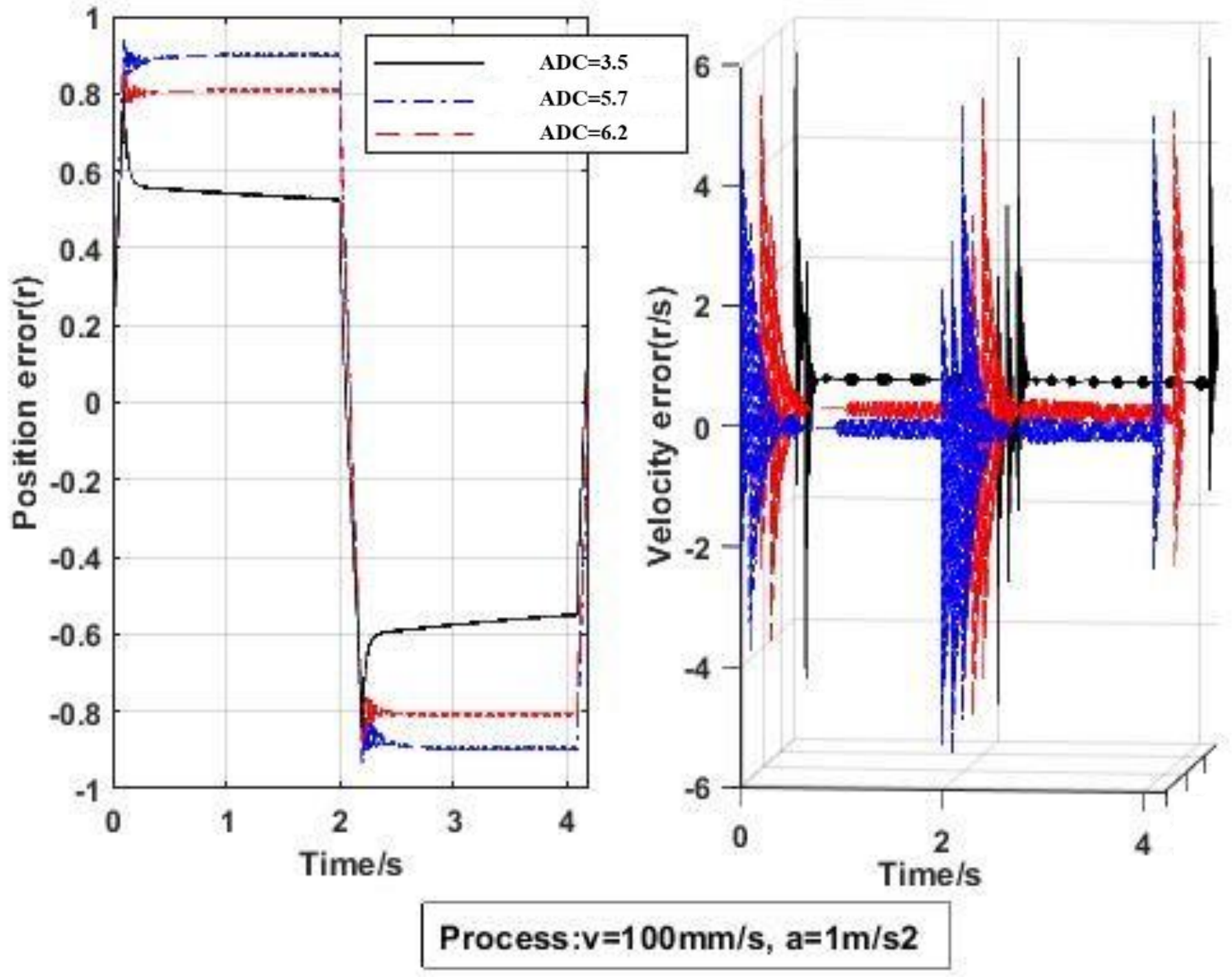

Figure 11 System performance when the acceleration is 1m/s$^2$

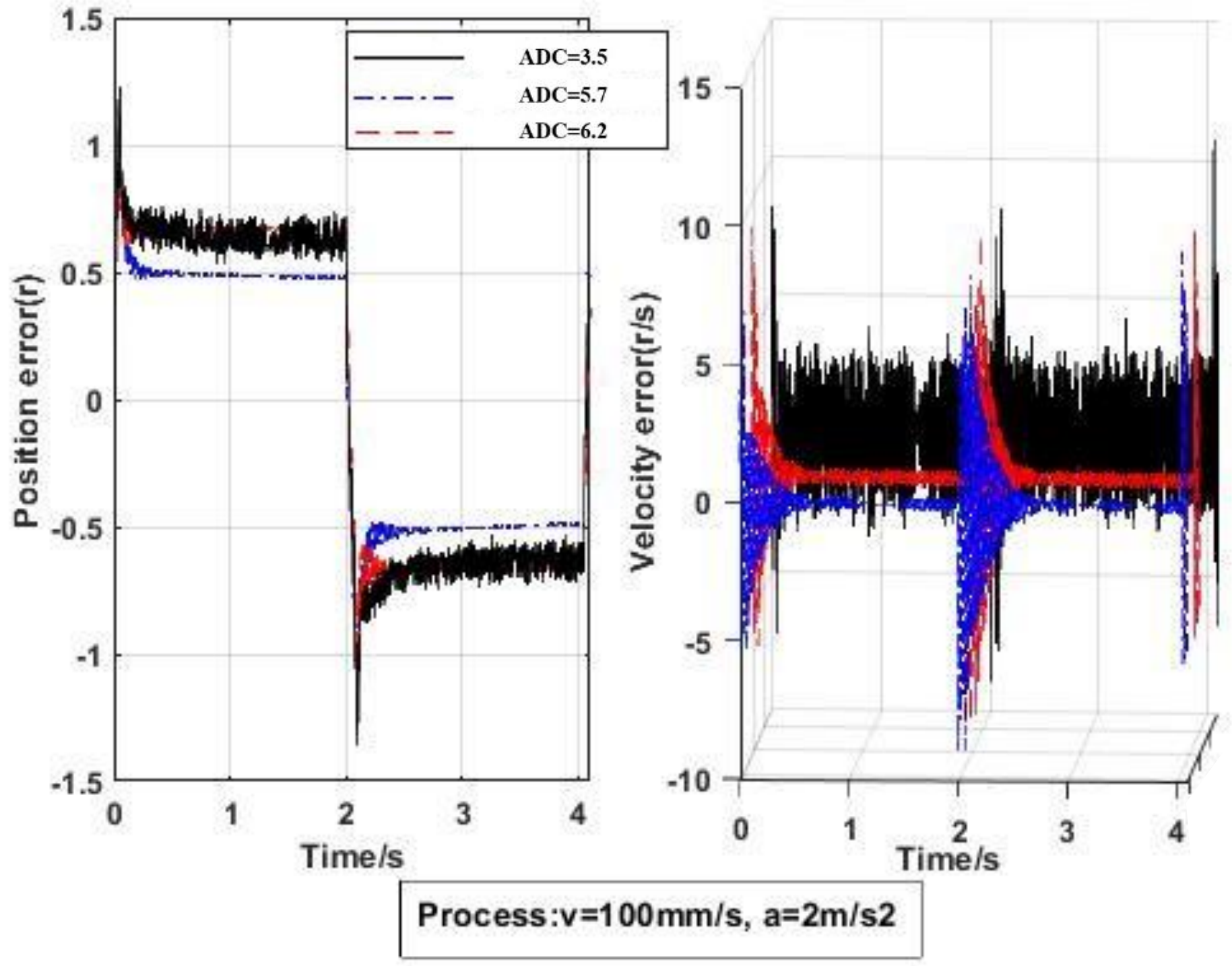

Figure 12 System performance when the acceleration is 2m/s$^2$

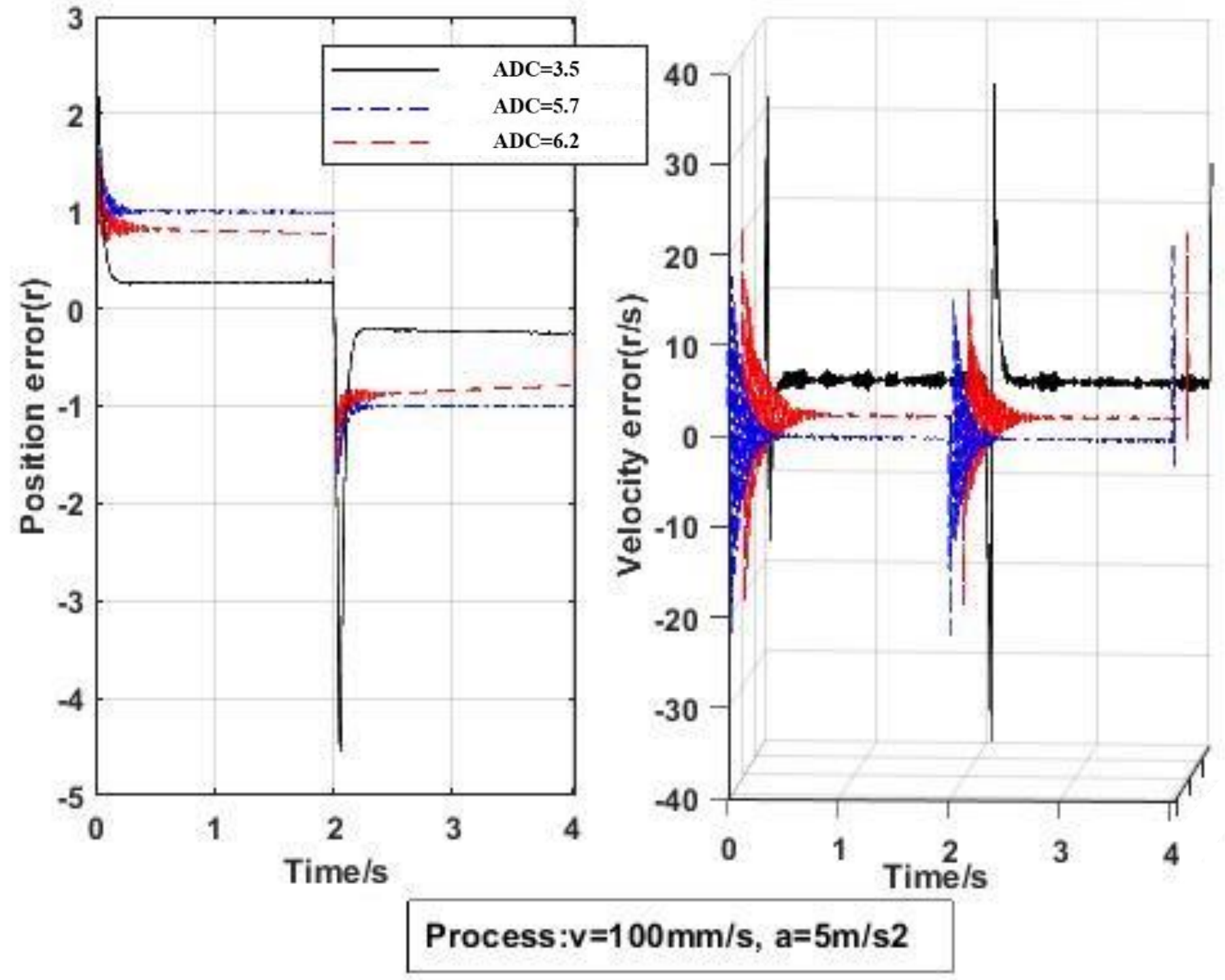

Figure 13 System performance when the acceleration is 5m/s[2]

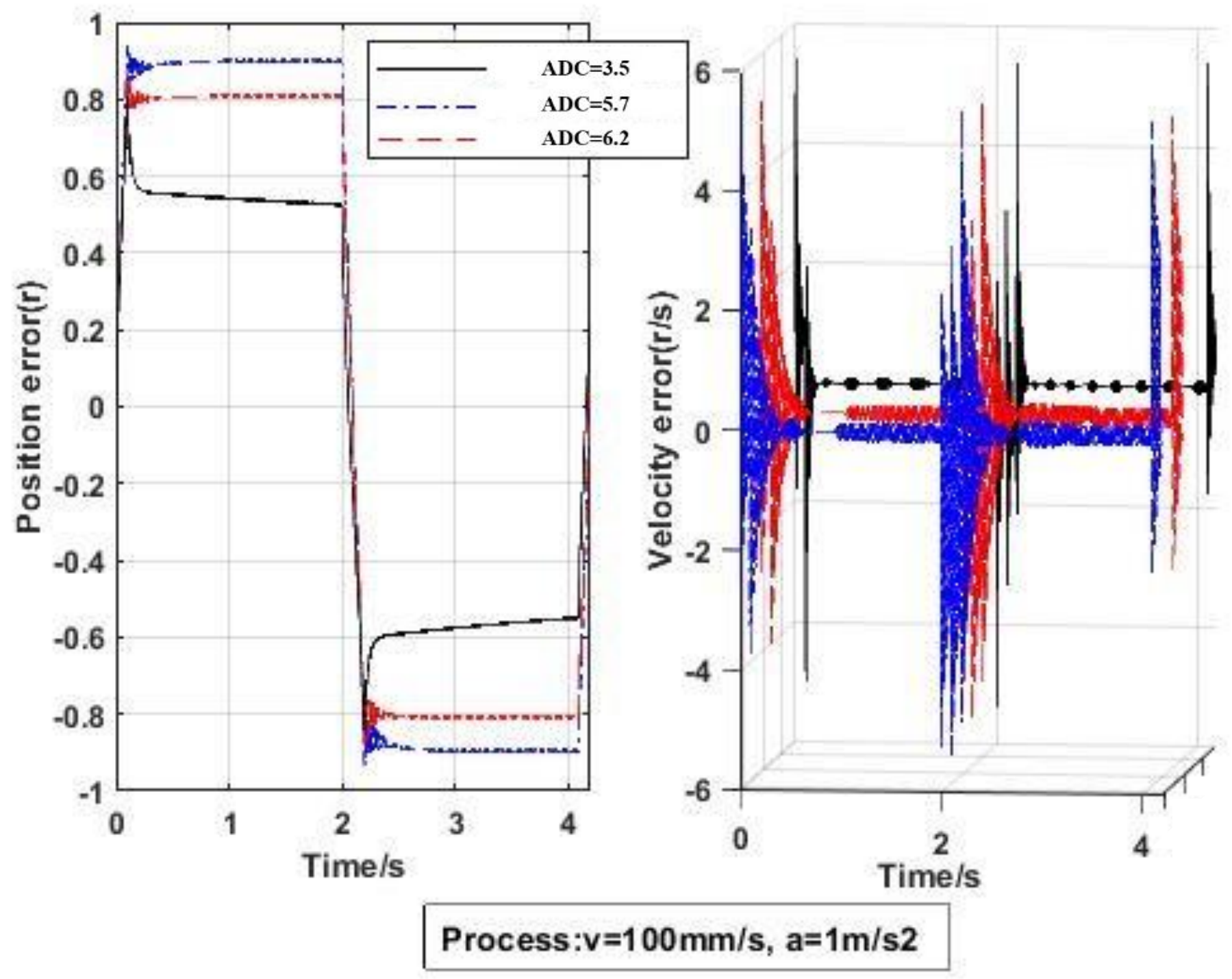

Figure 14 System performance when the max speed is 100mm/s

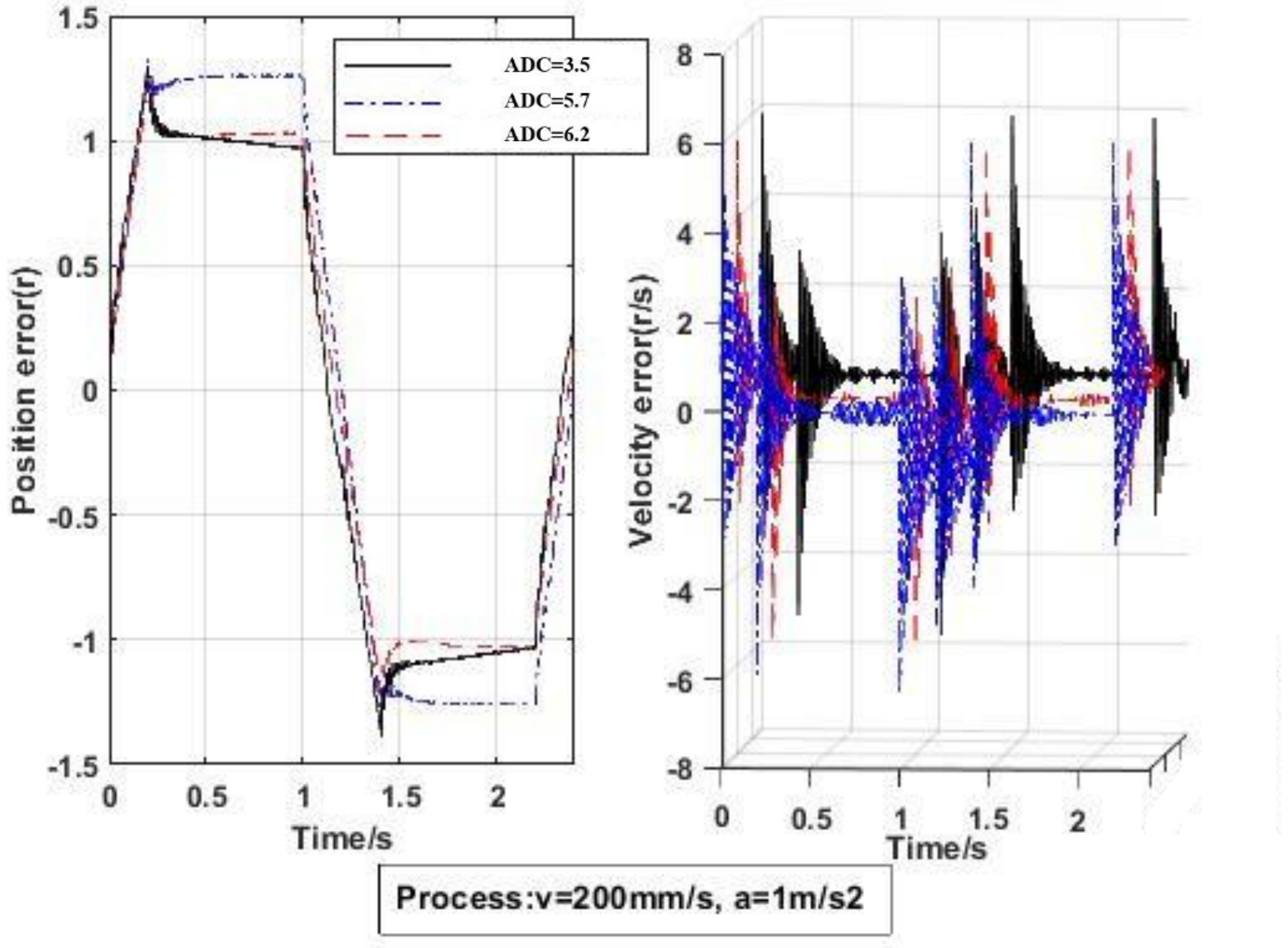

Figure 15 System performance when the max speed is 200mm/s

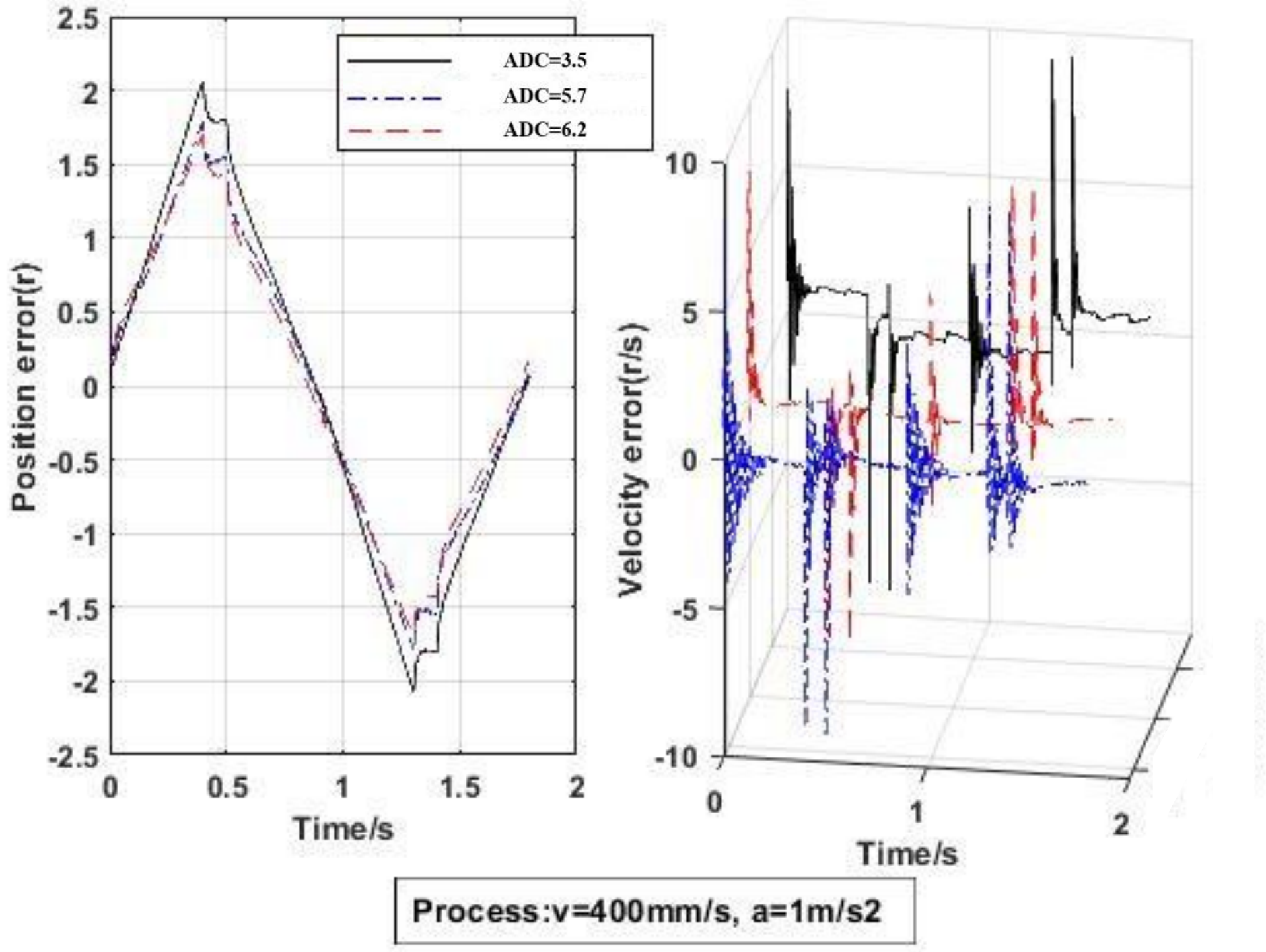

Figure 16 System performance when the max speed is 400mm/s

The overall performance of the system for different motors was obtained, as shown in Figure 17 and Figure 18. The horizontal axis represents the system's acceleration and deceleration capacity, while the vertical axis denotes the comprehensive dynamic performance index $W$ of the system. A smaller value indicates better dynamic performance of the system.

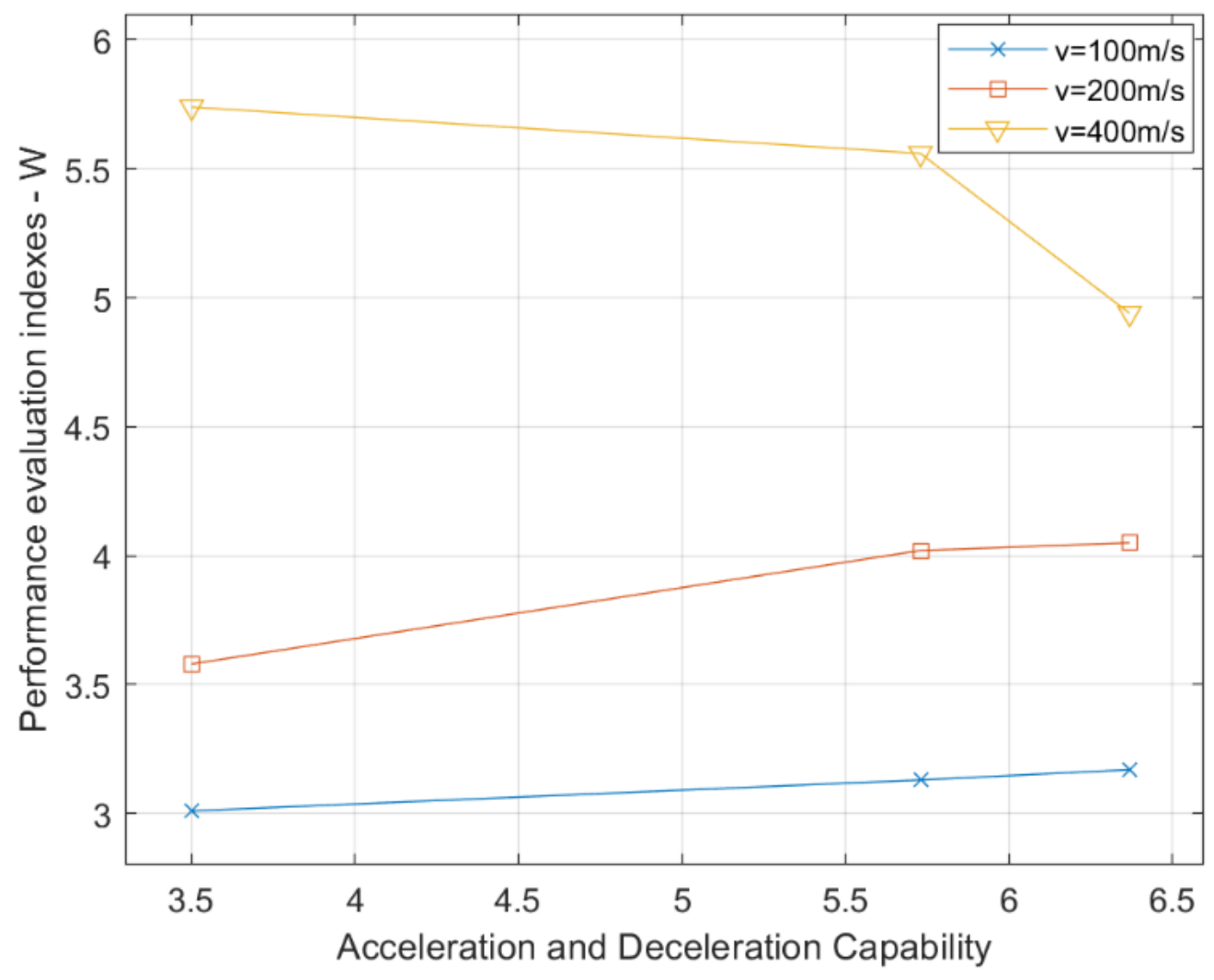

Figure 17 The influence of system acceleration and deceleration capacity on dynamic performance under different max speed

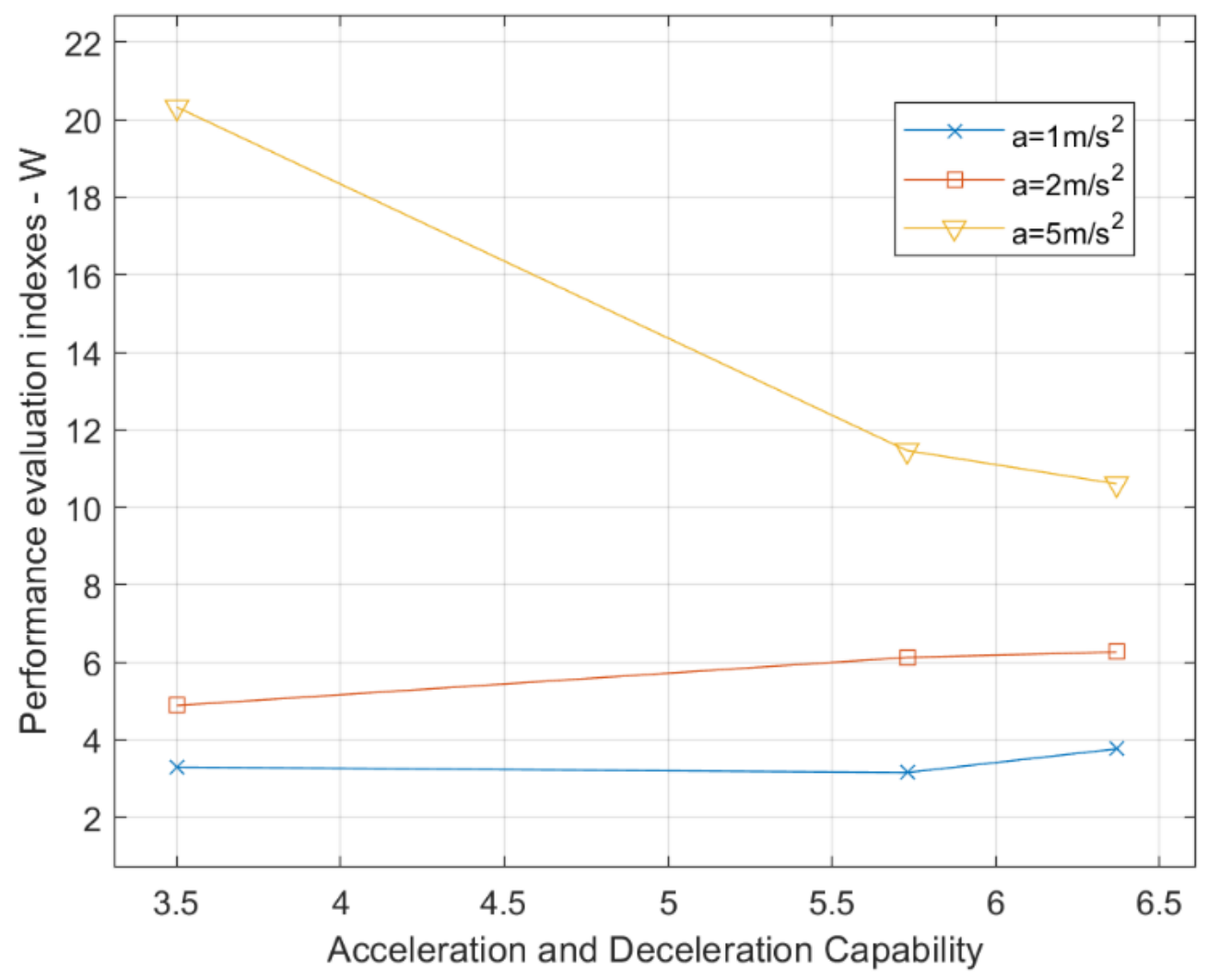

Figure 18 The influence of system acceleration and deceleration capacity on dynamic performance under different accelerations

As observed from Figure 17, when the speed of the motion process is relatively low, at 100 mm/s and 200 mm/s, the dynamic performance of the system deteriorates with the improvement of the system's acceleration and deceleration capacity. However, when the speed of the motion process is higher, at 400 mm/s, the dynamic performance of the system improves with the enhancement of the system's acceleration and deceleration capacity.

From Figure 18, it is evident that when the acceleration of the motion process is low, the dynamic performance of the system decreases with the improvement of the system's acceleration and deceleration capacity. Specifically, when the acceleration is 1 $m/s^2$ and 2 $m/s^2$, the decline in system dynamic performance ranges from 4.5% to 27% with the enhancement of acceleration and deceleration capacity. However, when the acceleration of the motion process is higher, at 5 $m/s^2$, the dynamic performance of the system improves with the enhancement of the system's acceleration and deceleration capacity.

This indicates that the system's acceleration and deceleration capacity should meet the actual requirements of the motion process. Both excessive and insufficient acceleration and deceleration capacity can lead to degradation in the system's dynamic performance. When acceleration and deceleration capacity are too low, the system lacks the ability to follow motion process instructions effectively, resulting in decreased dynamic performance. Conversely, excessive acceleration and deceleration capacity beyond the actual needs of the motion process can compromise system stability. The conclusions drawn from experimental and simulation results align with each other.

## 6 Conclusion

Currently, the comprehensive influence mechanisms of each subsystem on the overall dynamic performance of the feed system remain unclear. Existing design methods struggle to ascertain whether performance requirements can be met during the design phase, hindering the development of design approaches that directly focus on time domain precision. To address this issue, this paper proposes using acceleration and deceleration capability as a key parameter to characterize the electromechanical matching relationship. Furthermore, it considers the effects of both the control system and motion processes, allowing for an in-depth and comprehensive investigation of the combined impact of each subsystem on the time domain dynamic performance of the feed system. This study employs intelligent optimization algorithms to isolate the influence of the control system and investigates the comprehensive mechanism of the electromechanical system on the dynamic performance of the feed system under various motion processes. It yields the following key findings:

The acceleration and deceleration capacity of the system plays a pivotal role in determining the dynamic performance of the feed system. If the acceleration capacity is too small, the system may lack sufficient ability to follow movement instructions. When meeting the motion requirements of the feed system, the system performance decreases to a certain extent as the acceleration capacity increases. Because excessive acceleration and deceleration capacity beyond the actual needs of the motion process can compromise system stability. Therefore, to ensure dynamic performance and stability, the acceleration and deceleration capacity should be appropriately balanced – not too large or too small.

After decoupling the control system parameters, this study reveals the mechanisms by which different motion processes and electromechanical matching conditions affect the overall time domain dynamic performance of the feed system. This provides an important theoretical foundation for research into optimization design methods for CNC machine tool feed systems that directly target time domain dynamic precision.